\documentstyle[twocolumn,epsbox]{article}

\begin{document}

\twocolumn[%
\begin{flushright}
OCHA-SP-00-01
\end{flushright}
\title{
Characteristic length of random knotting for cylindrical self-avoiding polygons}
\author{
Miyuki K. Shimamura and Tetsuo Deguchi \\
\\
Department of Physics, Faculty of Science \\ 
and Graduate School of Humanities and Sciences, \\
Ochanomizu University \\
2-1-1 Ohtsuka, Bunkyo-ku, Tokyo 112-8610, Japan 
} 
\date{}
\maketitle 

\begin{abstract} 
We discuss the probability of random knotting for a model 
of self-avoiding polygons whose segments are given by cylinders 
of unit length with radius $r$.
We show numerically that the characteristic length of random knotting is roughly approximated by an exponential function of the chain thickness $r$.         
\end{abstract}

\vspace{3mm}
$\bf{keyword}$ \quad characteristic length, random knotting, self-avoiding polygons, \\
knot invariants,knots 
 
\vspace{1cm}]


\section{Introduction}

Knotted ring polymers such as knotted DNA molecules are 
synthesized in various experiments in chemistry and 
biology \cite{Dean,Shishido,Walba}.   
The question  on the topological constraint of the macromolecules 
was first formulated by Delbr{\"u}ck,  Frisch and Wasserman 
in the '60s. \cite{Delbruck,Frisch-Wasserman}  Then, 
topological properties of ring polymers 
have been studied through numerical simulations.
\cite{Vologodskii1,desCloizeaux,Michels,LeBret,Chen,Klenin,Janse,Koniaris,DeguchiJKTR,DeguchiRevE,Orlandini} Furthermore, the fractions of knotted species 
synthesized after the random cyclization of  
the circular DNAs are experimentally measured,  
and they give an experimental estimate on the effective diameter 
of the DNA molecules surrounded by the counter ions. 
\cite{Rybenkov,Shaw};    
it is estimated so that the experimental values of fractions of some knots   
are consistent with  the theoretical values of the fractions  obtained 
for the hedgehog model  \cite{Klenin} of self-avoiding polygons. The model   
gives self-avoiding polygons consisting of 
cylindrical segments, where the length and the radius of the cylindrical segments 
correspond to the Kuhn length and the effective radius, respectively.

\par 
Let us introduce the probability of random knotting. 
First, we assume an algorithm of constructing random polygons 
(or self-avoiding polygons) 
of $N$ polygonal nodes. 
Then, for a given knot $K$, we define knotting probability $P_K(N)$
by the probability of observing a random polygon 
(or a self-avoiding polygon) of the knot $K$. 
When $K$ is trivial ($K=0$), we denote it by $P_0(N)$.  
It has been found \cite{Michels,Koniaris} that the 
probability of unknot has 
a decreasing-exponential dependence on the nodes $N$: 
\begin{equation}  
P_0(N)= C_0 \exp(-N/N_0).  \label{unknot}
\end{equation}  
For the different  models of random polygons and self-avoiding polygons, 
knotting probabilities of some nontrivial knots 
have been evaluated numerically.  
\cite{Vologodskii1,Klenin,DeguchiJKTR,DeguchiRevE}
Through the simulations with the Vassiliev-type knot invariants, 
it has been found that for a given knot type $K$, 
the probability $P_K(N)$ as a function of the number $N$ of nodes 
can be  well described  by the following formula \cite{DeguchiJKTR,DeguchiRevE}
\begin{equation} 
P_K(N)=C_K \left({N \over {N_K}} \right)^{m(K)}
\exp\left(-N/N_K\right) .  
\label{general} 
\end{equation}
 From the numerical results, it is found that 
 the parameter $N_K$ should have almost the same 
value for any knot. \cite{DeguchiJKTR,DeguchiRevE}
The result is also consistent with the simulation  
of self-avoiding polygons on the cubic lattice. \cite{Orlandini} 
Thus, we may assume that $N_K$ should be  given by 
a constant $N_c$ for any knot $K$: $N_K = N_c$.   
We call $N_c$  the characteristic length of random knotting.

\par 
In this paper, we discuss how the  characteristic length 
$N_c$ depends on the excluded-volume parameter 
of self-avoiding polygons.    
We introduce a new method for constructing self-avoiding  polygons with 
cylindrical segments with radius $r$. 
The method is a variant of the dimerization algorithm.  
It seems that the algorithm is more efficient than that of the hedgehog model.   
Then, we study the characteristic length $N_c(r)$ 
as a function  of the cylinder radius $r$ 
for the self-avoiding polygons. 
We give  the estimates of the characteristic length 
by applying the formula (\ref{unknot}) 
to the numerical values  of the probability of unknot.  
Here the fitting parameters $C_0(r)$, $N_0(r)$ are determined by 
the least square method.

\par 
The numerical study on the characteristic length of random knotting  
as a function of the thickness parameter of polymer chains  
should be  important  in the study of 
knotted DNAs where the effective diameter can be changed according to the 
concentration of the counter ions. 
 Furthermore, it could be also important in other systems of polymer rings. 
The length and the radius of the cylindrical segments of the model 
correspond to the two fundamental 
parameters of polymer chains, i.e., the Kuhn length (the statistical length) 
and the stiffness (or the excluded-volume parameter),   
respectively. \cite{Grosberg-book}  
In association with  the numerical result of the rod-bead model \cite{Koniaris},  
 Nechaev and Grosberg have addressed      
a conjecture \cite{Nechaev} that 
the characteristic length $N_0$ as a function of some 
thickness parameter $d$ should be given by  the following:
\begin{equation}
N_0(d) = N_0(0) \exp(d/\ell) .  
\label{conj} 
\end{equation} 
Here $\ell$ denotes the Kuhn length.  
As we shall see later, however, the conjecture (\ref{conj}) does not hold for 
the rod-bead model.   On the other hand, the numerical simulation in this paper 
shows   that the exponential behavior 
(\ref{conj}) gives a rough approximation 
for  the new cylinder model of self-avoiding polygons.

\par 
The paper is organized as follows. In \S 2 
we introduce the cylinder model and explain the method of our numerical experiment 
of random knotting. 
In \S 3 we show the numerical data of our simulations and discuss possible relations 
between the characteristic length and the cylinder radius. We also discuss 
for the rod-bead model how the characteristic length depends on the bead radius. 
Finally, we give some discussion in \S 4.

\section{The methods of simulations} 
\subsection{New method for constructing self-avoiding polygons of cylindrical segments}

\par 
Let us explain the new method for 
constructing self-avoiding polygons  of cylindrical segments.   
In this method,  each segment of a polygon consists of a cylinder 
of unit length with radius $r$.  
The main structure of the algorithm is given by 
the following: (1) we generate a set of chains with cylindrical segments 
by the  dimerization method;  (2)  we construct polygons by 
connecting  two cylindrical self-avoiding chains 
with the concatenating method of Ref. \cite{Chen}. 
In this method,  we also calculate the statistical weight  
related to the probability of successful concatenation.

\par 
Let us define the condition of an overlap between a given pair of segments 
for our model of self-avoiding  polygons. 
First, we assume that there is no overlap between any pair of adjacent segments. 
Second, for a pair of two segments which 
are not next-neighboring to each other,  
we assume that the two segments have no overlap if and only if the distance 
between the central axes of the two cylinders 
is larger than $2r$. Here we have considered  
that a  central axis of a cylinder is given by 
the line segment between the centers of the upper and lower disks 
of the cylinder. 

\par 
In the step (1) of the algorithm, 
we construct a  chain with cylindrical segments randomly,  
and then we check whether there is an overlap 
or not for all unadjacent  pairs of cylinders of the chain. 
If there is an overlap, then we give up the chain and construct a new chain from
the beginning. If there is no overlap, then we keep the chain in the computer memory.

\par 
The new model of self-avoiding polygons produces free-joint rings 
with thickness parameter $r$, systematically. We can construct 
very long self-avoiding polygons with the cylinder radius $r$.  
It is known that the thickness of polymer chains 
plays an important role in the study 
of stiff polymers such as DNAs.  
\cite{Stiger,Brian,Yarmoia} We may assume 
that negatively charged  DNA molecules with surrounding counter ions 
can be approximated by  impermeable cylinders with the effective radius 
given by the screening effect. 
Thus,  the new algorithm can be applied to the study of knotted closed DNAs.

\subsection{The method for simulation of random knotting }

\par 
Let us describe the algorithm of our numerical simulation of random knotting. 
For a given number $N$ of polygonal nodes,  
we construct $M$ polygons. Here the number $M$ should be very large, such as  $M=10^4$. 
Then, for a fixed knot type $K$,  
we enumerate the number $M_K$ of such polygons out of the $M$ polygons 
that have the same set of
values of the knot invariants with the knot $K$.    
We estimate the probability  $P_K(N)$ of knot $K$ by 
the following 
\begin{equation}  
P_K(N) = {\frac {M_K} M} \, . 
\end{equation}

\par
We employ two knot invariants, the determinant $\Delta_K(-1)$ 
of knot  and the Vassiliev-type invariant $v_2(K)$ of the second degree, 
as the tool for detecting the knot type of a given polygon. 
 \cite{DeguchiPLA,Polyak} The values of the invariants 
for some typical knots are given in Table \ref{tab1}. 
\begin{table}
\caption{Values of the determinant of  knot $\mid \Delta_K(-1)\mid $  
and the second Vassiliev invariant $\it{v_2}$(K) for some simple knots.}
\label{tab1}
\begin{center}
\begin{tabular}{*{3}{c}}
  \hline 
   Knot K & $\mid \Delta_K(-1)\mid$  & $\it{v_2}$(K) \\
  \hline
    0    & 1 &  0  \\
   $3_1$ & 3 & -12 \\
   $4_1$ & 5 & 12  \\
   $5_1$ & 5 & -36 \\
   $5_2$ & 7 & -24 \\
   $3_1 \sharp 3_1$ & 9 & 24 \\
  \hline
\end{tabular}
\end{center}
\end{table}

\section{Characteristic lengths of random knotting }

\subsection{The exponential decay of the  probability of unknot}

\par 
Let us discuss the numerical results of our simulations.   
Employing the cylinder model introduced in \S 2,   
we construct self-avoiding polygons with   
 $N$ polygonal nodes where the cylinder radius  is given by $r$. 
For a given number $N$ and a given value of $r$, 
we construct four sets of $10^4$ polygons. ($M=4 \times 10^4$.) 
We consider for the polygonal nodes $N$ 
fourteen different numbers  from 20 to 150, 
and for the radius $r$ 10 different values  from 0.0 to 0.09. 
\begin{figure}[t]
\caption{
Probability $P_0(N)$ of unknot versus number $N$ of nodes for the
 cylinder model: the numerical estimates of $P_0(N)$ 
for $r=0.01$, 0.03 and 0.05 are shown 
by black circles, black triangles, and black diamonds, respectively, 
with error bars given by their standard deviations. Number $N$ of nodes are given by 
$10 j +1$ with $j=2,3, \ldots ,15$.}
\vspace{0.8cm}
\psbox[height=6.5cm,width=7.5cm]{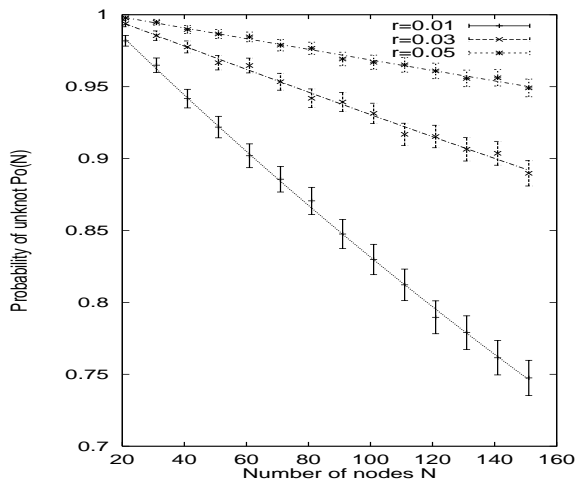}
\end{figure}

\par 
In Fig. 1, the estimates of the probability of unknot
for the cylinder model with $r=0.01$, 0.03, 0.05 are shown against the 
 number $N$ of polygonal nodes. 
The error bars are given by  the standard deviations.
For the cylinder model, we evaluate  
the variance of the values of the  probability $P_0(N)$ of unknot 
by taking into account both the statistical fluctuation of the number $M_K$ 
and that of the statistical weight \cite{Chen} 
appearing in the ring-dimerization procedure.

\par 
 From Fig. 1, we may confirm the exponential decay of $P_0(N)$ for the cylinder model.  
We note that the result is consistent with all the numerical simulations for other models
 such as  the Gaussian dynamical model \cite{Michels}, 
the Gaussian random polygon \cite{desCloizeaux,DeguchiRevE}, the hedgehog model \cite{Klenin}, 
and the rod-bead model \cite{Koniaris,DeguchiRevE}. 
The lines in Fig. 1 are theoretical curves given by the formula (\ref{unknot}).
The  estimates of the parameters  $C_0$ and $N_0(r)$ 
for the ten different values of the radius $r$ from $r=0$ to $r=0.9$ 
are given  in Table \ref{tab2}.
\begin{table}
\caption{Numerical estimates of the characteristic length $N_0(r)$ 
for the cylindrical model }
\label{tab2}
\begin{center}
\begin{tabular}{*{5}{c}}
  \hline
  radius $r$ & $C_0$ & $N_0(r)$ & $\chi^2$ \\
  \hline
  0.00 & 1.052$\pm$0.005 & (2.72$\pm$0.06) $\times$$10^2$ & 3\\
  0.01 & 1.028$\pm$0.004 & (4.72$\pm$0.14) $\times$$10^2$ & 1\\
  0.02 & 1.017$\pm$0.003 & (7.86$\pm$0.29) $\times$$10^2$ & 1  \\
  0.03 & 1.011$\pm$0.002 & (1.20$\pm$0.05) $\times$$10^3$ & 2\\
  0.04 & 1.007$\pm$0.002 & (1.88$\pm$0.10) $\times$$10^3$ & 2 \\
  0.05 & 1.006$\pm$0.001 & (2.64$\pm$0.16) $\times$$10^3$ & 2 \\
  0.06 & 1.004$\pm$0.001 & (3.99$\pm$0.31) $\times$$10^3$ & 1 \\
  0.07 & 1.003$\pm$0.001 & (5.56$\pm$0.48) $\times$$10^3$ & 1\\
  0.08 & 1.002$\pm$0.001 & (7.74$\pm$0.80) $\times$$10^3$ & 2 \\
  0.09 & 1.001$\pm$0.001 & (1.14$\pm$0.14) $\times$$10^4$ & 2\\
   \hline
\end{tabular}
\end{center}
\end{table}

\subsection{The characteristic length of the cylinder model}

\par 
A semilogarithmic plot of the numerical estimates of the characteristic length $N_0(r)$
is given in Fig. 2 against the cylinder radius $r$  for the ten different values  
from $r=0.0$ to $r=0.9$.     
\begin{figure}
\caption{Characteristic length $N_0(r)$ versus cylinder radius $r$ for the 
cylinder model. The numerical estimates of $N_0(r)$ listed in Table 2  
are depicted by black circles together 
with their errors. The fitting line of eq. (5)  
is determined  by the least square method with $\chi^2 = 42$.
}
\vspace{0.8cm}
\psbox[height=6.5cm,width=7.5cm]{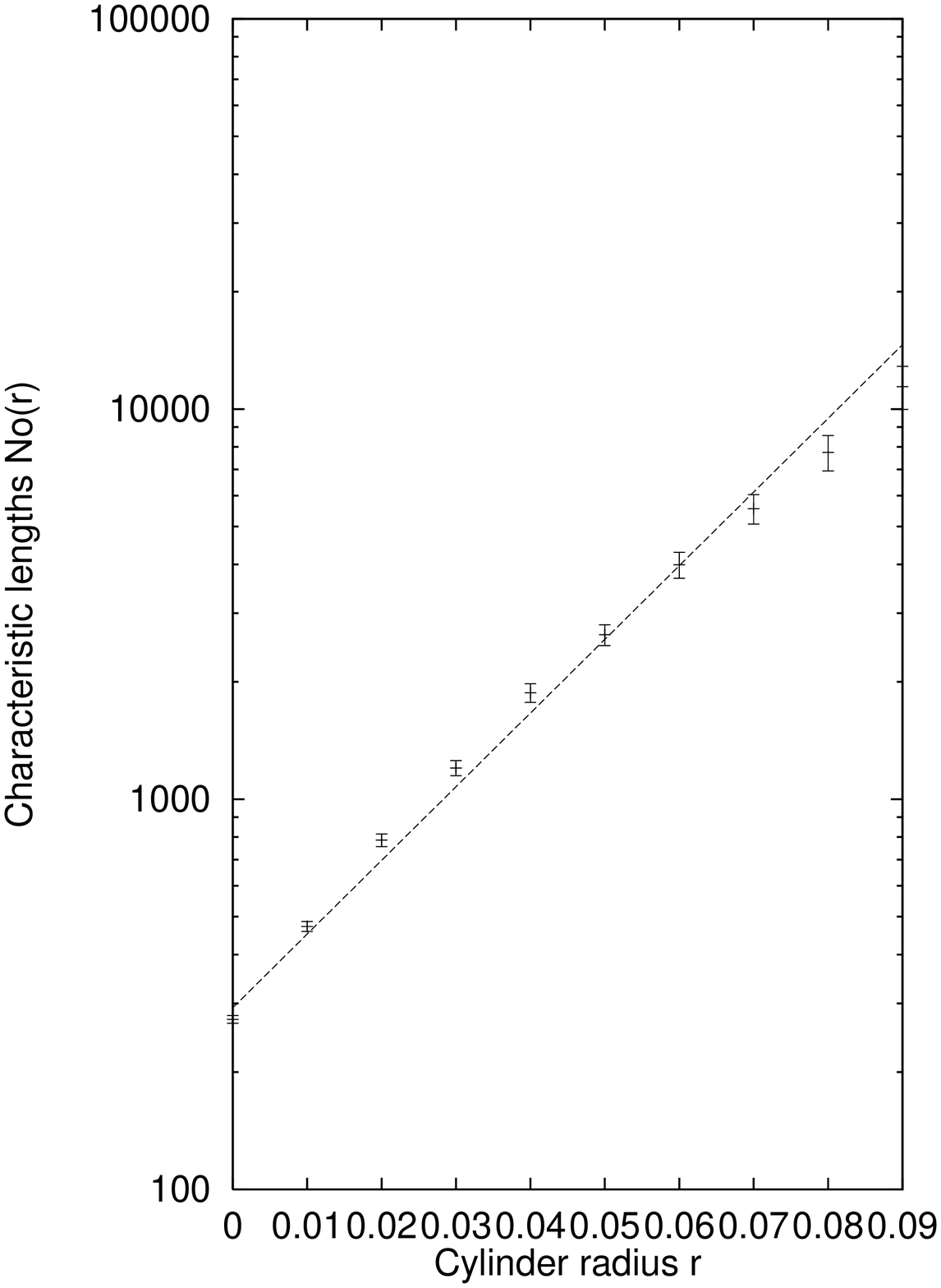}
\end{figure}
 From  Fig. 2,  we see that the characteristic length $N_0(r)$ as a function of 
the cylinder radius $r$ can be roughly 
approximated by  an increasing exponential function of $r$.  
Let us consider a fitting formula in the following 
\begin{equation}
N_0(r) =  N_{0}(0) \exp\left({\alpha} r \right) . 
\label{fit1} 
\end{equation}
Here, there are  two parameters to fit, i.e.,  $N_0(0)$ and $\alpha$. 
Applying the formula (\ref{fit1}) to the numerical estimates 
of the characteristic lengths, we have the fitting line drawn in Fig. 2. 
The estimates of the fitting parameters are given by 
$N_0(0)=292 \pm 5$ and $\alpha=43.5 \pm 0.6$. 
 From the viewpoint of the $\chi^2$ test, however, 
the curve does not fit well to the data. 
We note that $\chi^2 = 42$ while the number of data points is ten 
with the two fitting parameters. 
Let us now consider  another fitting formula 
\begin{equation}
N_0(r) =  N_{0}(0) \exp\left({\beta} \, r^{\nu}  \right), 
\label{stretch}
\end{equation}
where there are three parameters $N_0(0)$, $\beta$ and $\nu$. 
The new fitting curve is shown in Fig. 3 to the same numerical data points  
with Fig. 2. 
The best estimates of the  fitting parameters are given by 
 $N_0(0)=271 \pm 6$,  $\beta=29 \pm 2$  and $\nu = 0.85 \pm 0.02$.    
It seems that the  curve fits well to the data,   
although  the value of $\chi^2$ might be  a little too small: $\chi^2= 2.2$. 
Thus, we may conclude that for the cylinder model, 
the characteristic length of the probability of unknot 
as a function of the cylinder radius is given by the formula 
(\ref{stretch}). 
\begin{figure}[t]
\caption{Fitting curve of eq. (6)  to the data of $N_0(r)$ versus radius $r$ for the 
cylinder model.  The three fitting parameters of eq. (6) are determined by 
the least square method, which gives $\chi^2=2.2$.  
The data points and their error bars are 
given in Table 2. }
\vspace{0.8cm}
\psbox[height=6.5cm,width=7.5cm]{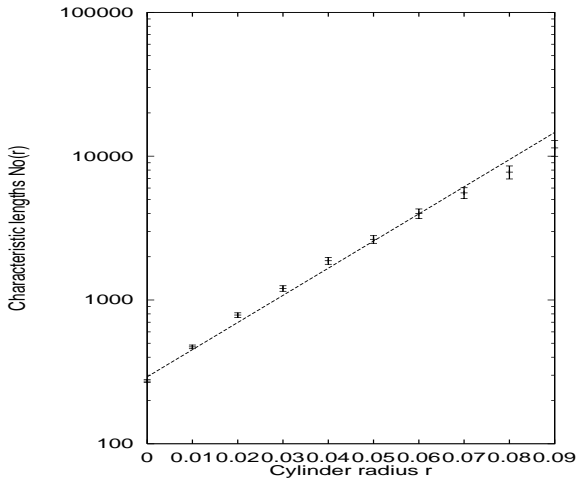}
\end{figure}

\par 
We have evaluated  the characteristic length $N_0(r)$  
for $r < 0.1$. 
For stiff chains such as DNAs, the radius $r$ can be 
rather small such as $r=0.003$. \cite{Grosberg-book} 
We thus consider that the range: $0.0 \le r < 0.1$ 
covers  most important cases. Furthermore, when $r \ge  0.1$, 
 numerical evaluation of the characteristic length becomes 
technically more difficult. We recall that even for the case of $r=0.09$,  
the estimates $N_0(r)$ 
is of the order of $10^4$, which is much larger than 
the polygonal nodes ($N < 160$) of our numerical simulations.

\subsection{The characteristic length  of the rod-bead model}

Let us now discuss the characteristic length  for the rod-bead 
model. Here we assume that the parameter $r$ denotes the bead radius.  
$(Fig. 4)$
\begin{figure}[t]
\caption{Two series of estimates of the characteristic length $N_0(r)$ 
are shown by black triangles for the rod-bead model 
 and by black circles for the cylinder model, respectively.  
The estimates for the cylinder model is given in Table 2, 
and those of  the rod-bead model in Table 3. 
The two estimates for the two models 
coincide for $r=0$.}
\vspace{0.8cm}
\psbox[height=6.5cm,width=7.5cm]{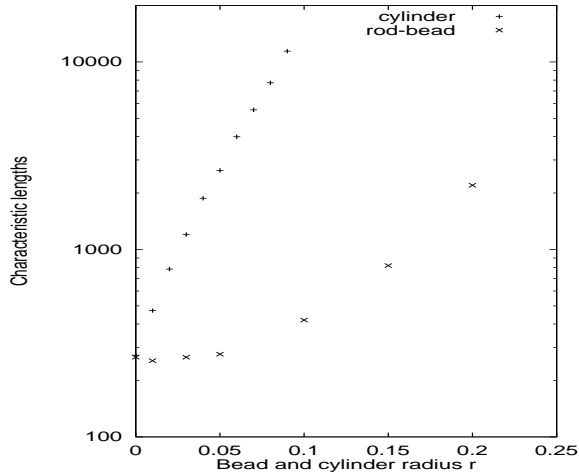}
\end{figure}
\begin{table}
\caption{Numerical estimates of the 
characteristic length of unknot for the rod-bead model.
The values  for $r=0.0$, 0.01, 0.03, 0.05, are calculated in this paper, 
 while those of  $r=0.1$, 0.15, 0.2 are given in Ref. [15], which 
are also consistent with Ref. [13]. All the data are obtained  
by constructing $10^4$ polygons ($M=10^4$). 
}
\begin{center}
\begin{tabular}{*{4}{c}}
  \hline
   radius $r$ & $C_0$ & N(K)\\
  \hline
   0.0 & 1.05$\pm$0.004 & (2.7$\pm$0.06) $\times$$10^2$\\
   0.01 & 1.07$\pm$0.01 & (2.6$\pm$0.08) $\times$$10^2$ \\
   0.03 & 1.06$\pm$0.01 & (2.6$\pm$0.09) $\times$$10^2$ \\
   0.05 & 1.05$\pm$0.01 & (2.8$\pm$0.1) $\times$$10^2$\\
   0.1  & 1.04$\pm$0.03 & (4.2$\pm$0.1) $\times$$10^2$ \\
   0.15 & 1.00$\pm$0.03 & (9.0$\pm$0.3) $\times$$10^2$ \\
   0.2 &  1.0$\pm$6.5 &(2.2$\pm$0.6) $\times$$10^3$  \\
  \hline 
\end{tabular}
\end{center}
\end{table}
In Fig. 4,  for the rod-bead model,  
 the numerical estimates of the characteristic length $N_0$ 
are plotted against the bead radius $r$.   
For an illustration, the estimates of the characteristic length $N_0$ 
for the cylinder model are 
also shown in Fig. 4 against the cylinder radius $r$. 
We note that when $r=0$, the characteristic lengths of the two models 
should coincide. The two algorithms are essentially  equivalent  
for $r=0$. From Fig. 4, we see that the numerical estimates 
of the two models really coincide for $r=0$, 
which gives a partial consistency check of our numerical simulations.

\par 
Nechaev and Grosberg discussed a conjecture that 
the characteristic length $N_0$ 
as a function of the ``chain thickness" $d$ should be given by 
an exponential form: $N_0(d) = N_0(0) \exp(27 d/\ell)$, where 
$\ell$ is the Kuhn length. \cite{Nechaev}     
It seems that their conjecture is based on the numerical result \cite{Koniaris} 
of the rod-bead model. 
However, we see from Fig. 4 that the 
characteristic length $N_0(r)$ for the rod-bead model 
is not given by an 
exponential function of the bead radius 
$r$  with $0.0 \le r \le 0.2$. 
Let us consider a part of the range such as $0.1 \le r \le 0.2$, and apply
for the rod-bead model an exponential approximation:    
$N_0(r) = N_0(0) \exp(\gamma \, r)$.   
Then, we find that the parameter $\gamma$ is roughly given by 15. 
Thus, for the rod-bead model, the conjecture of the exponential dependence 
could be  valid  only  for a limited range of the bead radius.
Furthermore, it seems that  
the estimate of the  coefficient $\gamma$ is not consistent with 
the conjectured value of 27.  

\section{Discussion}

 Finally, we discuss a possible connection of the cylinder model of this paper 
to the hedgehog model. Klenin $\it{et. al.}$ \cite{Klenin} 
studied the probability of unknot 
for the hedgehog model of ring polymers. 
The algorithm of the model is given by the following: 
(1) generating a set of vectors of unit length 
with a common origin (a ``hedgehog''); and (2) setting 
the minimal distance between any two unadjacent segments 
larger than the segment diameter.

\par 
It seems that the hedgehog model and the cylinder model introduced 
in this paper produce almost the same results 
 for the self-avoiding polygons with clylinder radius $r$.   
It is our conjecture that the two models should give the same results,  
although the two algorithms themselves are quite different.  
For the probability of unknot,  the two values obtained by the 
two different methods are rather close.  
However, it is not easy to see  whether  
they are exactly the same or not. 
We shall discuss  much more precise 
comparison  between  the two models 
in our future publications.

\newpage 
\par \noindent 
{\bf Figure captions} 

\par \noindent
 Fig. 1: 
Probability $P_0(N)$ of unknot versus number $N$ of nodes for the
 cylinder model: the numerical estimates of $P_0(N)$ 
for $r=0.01$, 0.03 and 0.05 are shown 
by black circles, black triangles, and black diamonds, respectively, 
with error bars given by their standard deviations. Number $N$ of nodes are given by 
$10 j +1$ with $j=2,3, \ldots ,15$.   


\vskip 0.6cm 
\par \noindent 
 Fig. 2:  
Characteristic length $N_0(r)$ versus cylinder radius $r$ for the 
cylinder model. The numerical estimates of $N_0(r)$ listed in Table 2  
are depicted by black circles together 
with their errors. The fitting line of eq. (5)  
is determined  by the least square method with $\chi^2 = 42$.

\vskip 0.6cm 
\par \noindent 
Fig. 3:  
Fitting curve of eq. (6)  to the data of $N_0(r)$ versus radius $r$ for the 
cylinder model.  The three fitting parameters of eq. (6) are determined by 
the least square method, which gives $\chi^2=2.2$.  
The data points and their error bars are 
given in Table 2. 

\vskip 0.6cm 
\par \noindent 
Fig. 4:  
Two series of estimates of the characteristic length $N_0(r)$ 
are shown by black triangles for the rod-bead model 
 and by black circles for the cylinder model, respectively.  
The estimates for the cylinder model is given in Table 2, 
and those of  the rod-bead model in Table 3. 
The two estimates for the two models 
coincide for $r=0$.


\end{document}